\documentclass[twocolumn,showpacs,aps,pra,amsmath,amssymb,floatfix]{revtex4}
\preprint{draft 1.5}

\usepackage{graphicx}
\usepackage{dcolumn}
\usepackage{bm}
\usepackage{amssymb}
\usepackage{setspace}

\newcommand{\Op}[1]{\mathsf{\hat{#1}}}

\begin{document}

\title{Enhancement of the formation of ultracold $^{85}$Rb$_2$ molecules due to resonant coupling}

\author{H. K. Pechkis}
\author{D. Wang}
\author{Y. Huang}
\altaffiliation{Present address: Akamai Physics, Inc., Las Cruces, NM 88005}
\author{E. E. Eyler}
\author{P. L. Gould}
\author{W. C. Stwalley}
\affiliation{Department of Physics, University of Connecticut, Storrs, CT 06269}

\author{C. P. Koch}
\affiliation{Institut f\"ur Theoretische Physik,
Freie Universit\"at Berlin,
Arnimallee 14, 14195 Berlin, Germany}

\date{\today}

\begin{abstract}

We have studied the effect of resonant electronic state coupling on the formation of ultracold
ground-state $^{85}$Rb$_2$.  Ultracold Rb$_2$ molecules are formed by photoassociation (PA) to a
coupled pair of $0_u^+$ states, $0_u^+(P_{1/2})$ and $0_u^+(P_{3/2})$, in the region below the
$5S+5P_{1/2}$ limit.  Subsequent radiative decay produces high vibrational levels of the ground
state, $X~^1\Sigma_g^+$.  The population distribution of these $X$ state vibrational levels is
monitored by resonance-enhanced two-photon ionization through the $2~^1\Sigma_u^+$ state.  We find
that the populations of vibrational levels $v''$=112$-$116 are far larger than can be accounted
for by the Franck-Condon factors for $0_u^+(P_{1/2}) \rightarrow X~^1\Sigma_g^+$ transitions with
the $0_u^+(P_{1/2})$ state treated as a single channel.  Further, the ground-state molecule
population exhibits oscillatory behavior as the PA laser is tuned through a succession of $0_u^+$
state vibrational levels.  Both of these effects are explained by a new calculation of transition
amplitudes that includes the resonant character of the spin-orbit coupling of the two $0_u^+$
states. The resulting enhancement of more deeply bound ground-state molecule formation will be
useful for future experiments on ultracold molecules.

\end{abstract}


\maketitle

\section{Introduction}

\begin{figure}
\centerline{\includegraphics[width=0.98\linewidth]{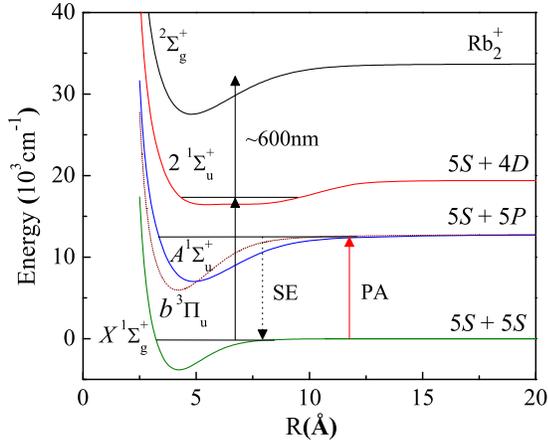}}
\caption{(color online) Selected potential curves for Rb$_2$ from
  Ref. \cite{ParkJMS01}, showing our scheme for forming ultracold $^{85}$Rb$_2$ molecules in the
$X^1\Sigma_g^+$ state.  Photoassociation (PA) is used to excite high-$v$ levels of the coupled
$0_u^+$ states, $0_u^+(P_{1/2})$ and $0_u^+(P_{3/2})$, converging to the $5S+5P_{1/2}$ and $5S+5P_{3/2}$ atomic limits.  These states correlate at short range to the $A\ ^1\Sigma_u^+$ and $b\ ^3\Pi_u$ states.  Spontaneous emission (SE) from these levels populates the ground $X$ state. Pulsed laser light at ~600 nm is used for ionization detection.}
\label{fig:potentials}
\end{figure}

In recent years, the technique of photoassociation (PA) has become a powerful method for the
formation and study of ultracold molecules produced from ultracold atoms \cite{Jones06,Doyle04}.
Although PA can form molecules only in electronically excited states, subsequent radiative decay
can produce ground-state molecules, mainly in their highest vibrational levels.  To enhance the
formation of ground-state molecules in levels well below the dissociation limit, it would be ideal
to photoassociate to an excited state that has a bimodal vibrational probability distribution,
with one peak at long range to enhance the photoassociation rate, and another at shorter range to
enhance radiative decay into deeply-bound levels of the ground electronic state.  Theoretical and
experimental work on ultracold Cs$_2$ \cite{DionPRL01,Pichler06} has shown that such states can exist
because of resonant spin-orbit coupling between the vibrational series of the two 0$_u^+$ states
that converge to the ($6S+6P_{1/2}$) and ($6S+6P_{3/2}$) fine-structure limits.  In Ref.
\cite{DionPRL01} the formation of more deeply-bound vibrational levels of Cs$_2$ is greatly
enhanced when this coupling is at its maximum, and it was proposed that this approach might be
equally applicable for other alkali dimers. There have been several additional studies of the
effects of the resonant coupling phenomenon on photoassociation spectroscopy and efficient
formation of ultracold molecules \cite{DulieuJOSA03,AmiotPRL99,BergemanJPhysB06,JelassiPRA06b}.

In Rb$_2$ the spin-orbit perturbations between the pair of 0$_u^+$ states converging to
($5S+5P_{1/2}$) and ($5S+5P_{3/2}$) have been studied in detail
\cite{BergemanJPhysB06,JelassiPRA06b}, and agreement between theory and experiment for the 0$_u^+$
energy levels is very good.  However, the effects on ground-state Rb$_2$ formation have not been
investigated until quite recently.  In 2006, anomalous vibrational distributions were reported that could not be fully explained using Franck-Condon factors calculated from adiabatic potentials \cite{Huang06a}.  A recently completed comparative study of molecule formation in $^{85}$Rb and $^{87}$Rb attributes most of the differences between the isotopes to resonant coupling \cite{Fioretti07}.  In this paper we report a detailed experimental and theoretical study of resonant enhancement of ground-state $^{85}$Rb$_2$ molecule formation, with an emphasis on the energy dependence for a range of vibrational levels of the 0$_u^+$ state.  As shown in Fig. \ref{fig:potentials}, we photoassociate pairs of colliding $^{85}$Rb atoms into high molecular vibrational levels of the two $0_u^+$ states in the region just below the $5S+5P_{1/2}$ asymptote.  Some of these molecules decay radiatively to the $X\,^1\Sigma_g^+$ state, forming either bound molecules in high-$v''$
levels or unbound pairs of atoms.  The bound ground-state molecules are detected
state-selectively by two-photon ionization via the $2\,^1\Sigma_u^+$ state.

We focus particularly on the variation of molecule formation rates for the vibrational levels $v''=112-116$ as a function of the vibrational level $v'$ of the coupled 0$_u^+$ states to which PA takes place.  A coupled-channels model is used to predict Franck-Condon factors and transition moments for the 0$_u^+ \rightarrow X\,^1\Sigma_g^+$ radiative decay process, as well as the detection process. These results are compared with measurements of the relative formation rates after carefully normalizing them to remove dependencies on the PA rates and extraneous experimental factors.

\section{Theory}
\label{sec:theory}

The coupled 0$_u^+(P_{1/2})$ and 0$_u^+(P_{3/2})$ states arise due to spin-orbit coupling of the
short-range $A^1\Sigma_u^+$ and $b^3\Pi_u$ states.  Dipole-allowed radiative decay to the ground
$X^1\Sigma_g^+$ state is possible only for the portion of the 0$_u^+$ wavefunction having
$A^1\Sigma_u^+$ character.  Vibrational eigenfunctions of the $0_u^+$ excited states, the
$X\,^1\Sigma_g^+$ electronic ground state and the $2\,^1\Sigma_u^+$ excited state are obtained in
order to calculate Franck-Condon factors needed to predict  the relative efficiency of formation
and detection of ground-state  molecules in various vibrational levels $v''$.

\subsection{Model for the molecule formation process}
\label{subsec:model}

In the diabatic basis, the Hamiltonian describing nuclear motion in the electronically excited 0$_u^+(P_{1/2})$ and 0$_u^+(P_{3/2})$ states
reads
\begin{equation}
  \label{eq:Hexc}
      \Op{H}_e =
    \begin{pmatrix}
      \Op{T} + V_{A^1\Sigma_u^+}({R}) & W^{off}_\mathrm{SO}({R}) \\[2ex]
      W^{off}_\mathrm{SO}({R}) &
      \Op{T} + V_{b^3\Pi_u}({R}) - W^{dia}_\mathrm{SO}({R})
    \end{pmatrix} \, ,
\end{equation}
where ${R}$ denotes the internuclear coordinate, $\Op{T}$ the
corresponding nuclear kinetic energy operator, $V_i({R})$ the
respective potentials, and $W^j_\mathrm{SO}({R})$ the diagonal and
off-diagonal spin-orbit coupling functions.  The hyperfine interaction
in the excited state can be neglected because PA detunings of several
wavenumbers are considered.  We are interested in the formation of
ground-state molecules with binding energies $E_{v''}>3\,$cm$^{-1}$, much larger than the ground-state hyperfine splitting of 0.1 cm$^{-1}$ for $^{85}$Rb.
Therefore, the hyperfine interaction in the ground state can also be
neglected, and the $X\,^1\Sigma_g^+$ ground state is hence described
in a single channel picture.  The $2\,^1\Sigma_u^+$ state can also be
treated as a single channel.  The corresponding Hamiltonians are
\begin{equation}
  \label{eq:Hg}
  \Op{H}_i = \Op{T} + V_{i}({R})\,
\end{equation}
with $i=X\,^1\Sigma_g^+,2\,^1\Sigma_u^+$.

The potentials $V_i({R})$ for the  $A\,^1\Sigma_u^+$ and $b\,^3\Pi_u$ states and the spin-orbit coupling functions $W^j_\mathrm{SO}({R})$ are taken from Ref. \cite{BergemanJPhysB06}. The ground-state potential is obtained by connecting the short range potential curve of Ref.~\cite{SetoJCP00} to the long-range expression $C_6/{R}^6+C_8/{R}^8+C_{10}/{R}^{10}$ with the $C_j$ coefficients found in Ref.~\cite{MartePRL02}.  The repulsive wall of the ground-state potential is adjusted to yield a singlet scattering length of 2400$\,a_0$ for $^{85}$Rb \cite{vanKempenPRL02}. The potential curve for the $2\,^1\Sigma_u^+$ state is taken from
Ref.~\cite{ParkJMS01}.

The Hamiltonians, Eqs.~(\ref{eq:Hexc}-\ref{eq:Hg}), are represented on a coordinate space grid employing a mapped grid method~\cite{SlavaJCP99,WillnerJCP04}.  Vibrational energy levels and wavefunctions are obtained by diagonalization, and Franck-Condon factors are calculated by evaluating the appropriate integrals.  For the $0_u^+ \rightarrow X\,^1\Sigma_g^+$ transitions, Franck-Condon factors were compared to transition dipole matrix elements where the transition dipole was included in the integrals as a function of internuclear coordinate~\cite{BeucPRA07}.  No significant differences were found in the relative transition amplitudes, consistent with the expected nearly constant long-range behavior of the transition dipole for the range of PA detunings considered
here.

\begin{figure}[bt]
  \centering
  \includegraphics[width=0.9\linewidth]{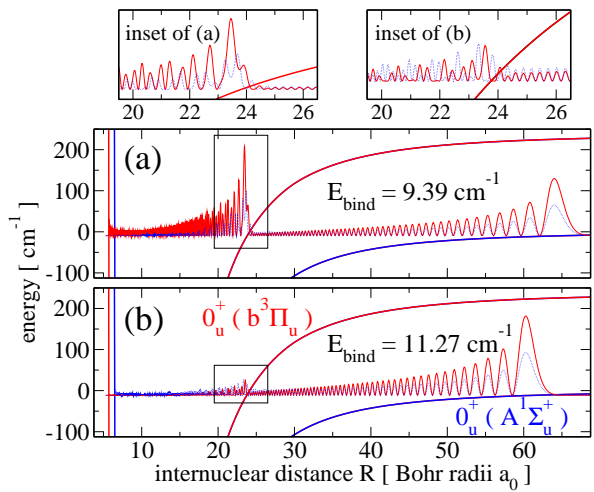}
  \caption{(color online)
    The squares of the vibrational eigenfunctions of the $0_u^+$ coupled states for
    binding energies of 9.39$\,$cm$^{-1}$ and
    11.27$\,$cm$^{-1}$ (red solid line: triplet component, blue
    dotted line: singlet component).
    The vertical position of the square of each wavefunction corresponds to its
    binding energy.  The potentials (thick solid lines) are drawn in
    order to illustrate how the structure of the coupled wavefunctions
    can be understood in terms of classical turning points.}
  \label{fig:wvfct}
\end{figure}

\subsection{Resonant enhancement of ground-state molecule formation
  with binding energies of $2-20$ cm$^{-1}$}
\label{subsec:theores}

It is well-known~\cite{AmiotPRL99,SlavaJCP99,SlavaPRA00} that the $0_u^+(P_{1/2})$ and $0_u^+(P_{3/2})$ states of Rb$_2$ constitute a classic example of strongly coupled electronic states.  This implies that throughout the vibrational spectrum, some of the $0_u^+$ eigenfunctions are strongly perturbed by the resonant coupling while others show fairly regular behavior. Fig.~\ref{fig:wvfct} compares strongly perturbed and regular vibrational eigenfunctions with binding energies of 9.39$\,$cm$^{-1}$
and 11.27$\,$cm$^{-1}$, respectively.

The spectral positions at which the two $0_u^+$ components are most
strongly coupled can be readily identified, e.g. by inspection of the
rotational constants $B(v)$ (cf. Fig.~6 of
Ref.~\cite{BergemanJPhysB06}).  The behavior of the square of the
wavefunction of Fig.~\ref{fig:wvfct}~(a) arises from the resonant
coupling of the eigenfunctions of the two $0_u^+$ potentials, showing
two classical turning points for each of the two potentials. The enhanced
probability amplitude at the outer classical turning point of the
upper $0_u^+(P_{3/2})$ potential, at about $R=23.5$~$a_0$, is expected
to lead to large
Franck-Condon factors with vibrational levels in the electronic ground
state. The variation of the square of the wavefunction in the vicinity of this
classical turning point is emphasized in the inset of
Fig.~\ref{fig:wvfct}~(a).  By contrast, the square of the vibrational wavefunction
in Fig.~\ref{fig:wvfct}~(b) shows almost regular behavior with only a
small perturbation visible around  the outer classical turning point
of the $0_u^+(P_{3/2})$ potential.  Note that the scale of the inset
of  Fig.~\ref{fig:wvfct}~(a) is three times the scale of the inset of
Fig.~\ref{fig:wvfct}~(b).

\begin{figure}[bt]
  \centering
  \includegraphics[width=0.9\linewidth]{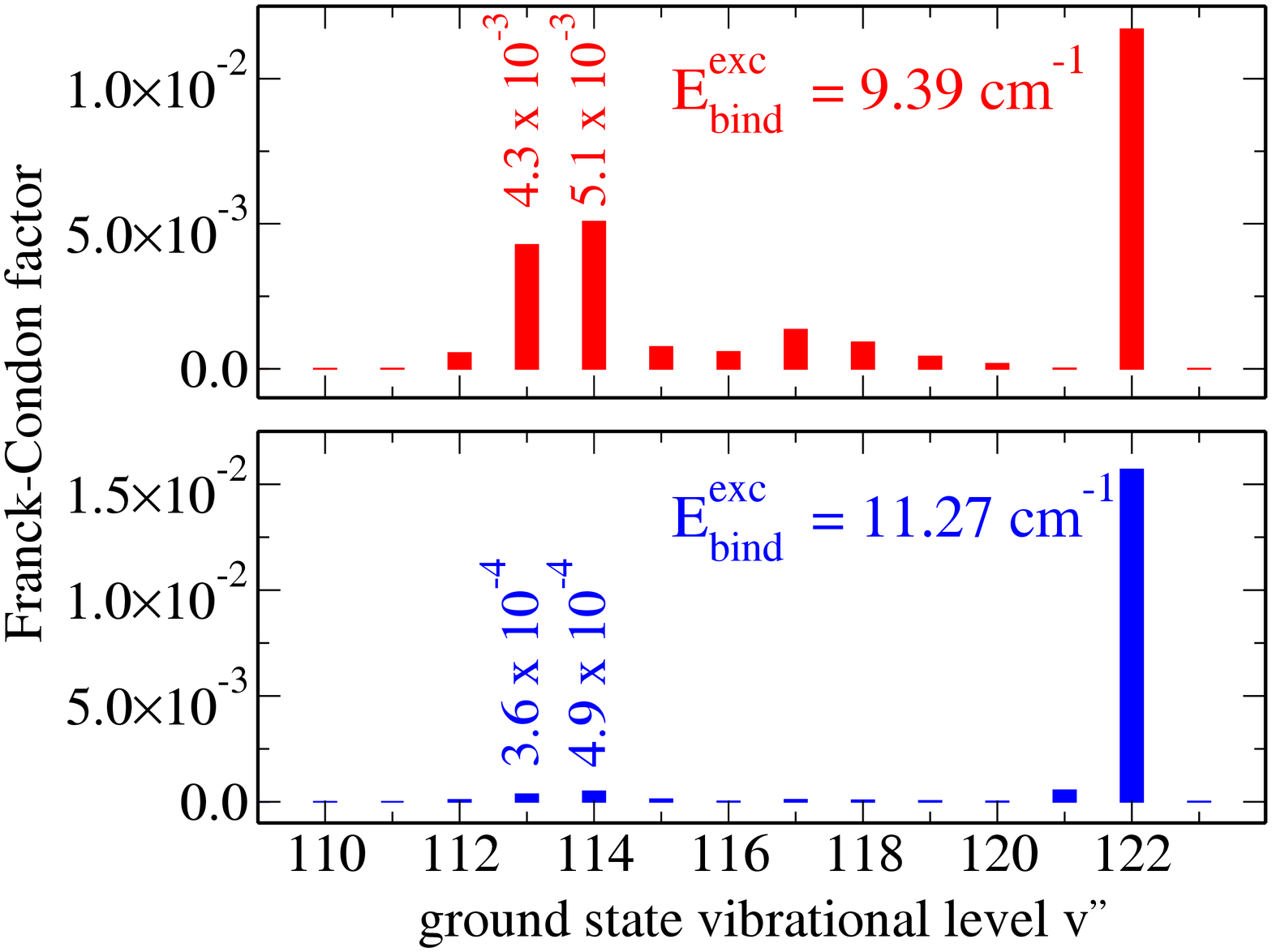}
  \caption{(color online)
    Franck-Condon factors between ground-state vibrational levels
    $v''$ and the $0_u^+$ vibrational levels with
    binding energies of 9.39 cm$^{-1}$ (top) and
    11.27 cm$^{-1}$ (bottom).  Enhancement of Franck-Condon factors by an order of
    magnitude due to the resonant coupling is observed for the
    $X\,^1\Sigma_g^+$ levels $v''=113$ ($E_\mathrm{bind}=6.9$ cm$^{-1}$)
    and $v''=114$ ($E_\mathrm{bind}=5.0$ cm$^{-1}$).}
  \label{fig:FCF}
\end{figure}

The Franck-Condon factors for each of the two $0_u^+$ eigenfunctions of Fig.~\ref{fig:wvfct} with
the highest vibrational levels of the $X^1\Sigma_g^+$ state are presented in Fig.~\ref{fig:FCF}.
For regular vibrational wavefunctions, one would expect larger $0_u^+$ binding energies
(corresponding to larger PA detunings) to result in larger Franck-Condon factors with the
ground-state vibrational levels $v''$.  The Franck-Condon factor for the second to last
$X^1\Sigma_g^+$ level ($v''=122$) is indeed slightly larger for the $0_u^+$ level bound by 11.27
cm$^{-1}$ than for the level at 9.39$\,$cm$^{-1}$. The probability density of the level $v''=122$
has its outermost peak at about $R=60$~$a_0$, just like the $0_u^+$ level with
$E_\mathrm{bind}=11.27$ cm$^{-1}$ (cf. Fig.~\ref{fig:wvfct}), while the outermost peak of the
$0_u^+$ level with $E_\mathrm{bind}=9.39$~cm$^{-1}$ occurs at larger distances, about
$R=64$~$a_0$.  The main contribution to the Franck-Condon factors with $v''=122$ can therefore be
attributed to the long-range portions of the wavefunctions.  These long-range maxima are also
responsible for efficient photoassociation.

\begin{figure}
\centering
\includegraphics[width=0.9\linewidth]{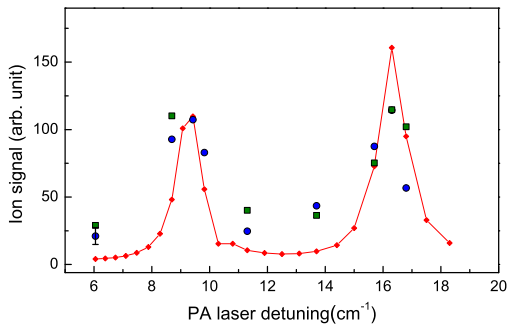}
\caption{(color online) Solid curve with red diamonds: Sums of calculated Franck-Condon factors from each of 25 vibrational levels of the coupled $0_u^+$ states to ground-state levels with $v''=112-116$, revealing strong oscillations in ground-state molecule formation.  The $0_u^+$ level energies are shown as a function of PA laser detuning below the $5S + 5P_{1/2}$ limit for comparison with experimental data.  Crosses: Experimental data set 1.  Circles: Experimental data set 2, taken on a different day under improved conditions as explained in Section~\ref{sec:results}. The typical systematic error bar shown is based on upper and lower bounds for systematic corrections as explained in the text.}
\label{fig:FCFoscillations}
\end{figure}

In addition to this long-range contribution to the Franck-Condon factors, the enhanced probability
density at intermediate distances (about $R=23.5$~$a_0$ in Fig.~\ref{fig:wvfct}) due to the
resonant coupling leads to significant Franck-Condon factors with the ground-state levels
$v''=112-119$, in particular with $v''=113$ and 114. Since the $0_u^+$ eigenfunction with
$E_\mathrm{bind}=9.39\,$cm$^{-1}$ is strongly perturbed by the resonant coupling, its
Franck-Condon factors with $v''=113$ and 114 are larger by one order of magnitude than those of
the $0_u^+$ eigenfunction with $E_\mathrm{bind}=11.27$~cm$^{-1}$. The enhancement of Franck-Condon
factors for the levels $v''=112-119$ due to strong perturbation of some of the $0_u^+$ vibrational
wavefunctions will result in efficient formation of ground-state molecules in these levels which
can be probed by varying the PA detuning.  This is the first important conclusion to be drawn from
Figs.~\ref{fig:wvfct} and \ref{fig:FCF}. The effect is obvious in the solid curve and red diamonds
of Fig.~\ref{fig:FCFoscillations}, which shows the integrated Franck-Condon factors from each
$0_u^+$ vibrational level to ground-state levels with $v''=112-116$ (the levels $v''=117-119$ are
omitted because they cannot be efficiently detected in our experiment for reasons explained
below).  Here the integration over a small range of $v''$ smooths out the nodal oscillations that
would be seen in the Franck-Condon factors to any single ground-state level $v''$, leaving only
the overall structure due to resonant coupling of the $0_u^+$ states.

Moreover, as seen in Fig. \ref{fig:FCF}, the Franck-Condon factors of the coupled-state $0_u^+$
eigenfunction of binding energy $E_\mathrm{bind}=9.39$~cm$^{-1}$ with the ground state levels
$v''=113$ and 114 differ only by a factor of about 2.7 and 2.3, respectively, from the
Franck-Condon factor with the second to last level,  $v''=122$.  While the levels $v''=113$ and
114 are highly excited, their binding energies are much larger than that of $v''=122$
($E_\mathrm{bind}=9.16\,$cm$^{-1}$ and 6.86$\,$cm$^{-1}$, respectively, compared to $\approx
10^{-2}\,$cm$^{-1}$ \cite{v=122}).  The outermost peaks
of the corresponding vibrational wavefunctions occur at about 21.4~$a_0$ and 22.2~$a_0$.  This
leads to the second important conclusion: Molecule formation in these more strongly bound
ground-state levels is almost as efficient as in the very last levels just below the dissociation
threshold.  This appears to resolve a puzzle raised in Ref. \cite{Huang06a}, where it was observed
that the vibrational populations of levels near $v''=112$ are much higher than can be explained by
Franck-Condon factors calculated using the uncoupled $0_u^+(P_{1/2})$ potential curve.

\section{Experiment}

The experimental setup has previously been described in detail in Ref. \cite{Huang06a}.  It
consists of a $^{85}$Rb MOT, which is loaded to a density of approximately 10$^{11}$ cm$^{-3}$
using a ``dark-SPOT'' \cite{Ketterle93} configuration.  The estimated MOT temperature is ~200
$\mu$K.  A cw tunable Ti:Sapphire laser (Coherent 899-29) with an output power of more than 400 mW
is focused into the MOT for photoassociation.  The photoassociated ultracold molecules
spontaneously decay to the ground state, $X~^1\Sigma_g^+$, which is subsequently detected by
resonance-enhanced two-photon ionization using a pulsed dye laser (Continuum ND 6000) pumped by a
532 nm Nd:YAG laser at a 10 Hz repetition rate.  The detection laser beam has a 1 mm diameter and
a typical pulse energy of 3 mJ with a pulse duration of 7 ns.  The molecular ions that it forms
are accelerated into a channeltron detector and discriminated from atomic ions by their time of
flight.

With the PA laser frequency fixed on a resonance, the pulsed-detection laser can be scanned in the
range of 601-603 nm to obtain the vibrational spectrum of the ground-state molecules.  Since the
theory of resonant coupling predicts an oscillatory behavior of the number of formed molecules as
a function of PA laser detuning, we repeat the experiment for nine different PA laser detunings,
each time obtaining a detection laser scan.  To compare more clearly the effect of resonant
coupling for these different PA laser detunings, we normalize each detection spectrum by the
fractional PA trap loss measured at the same PA laser detuning.  More details regarding the
normalization are given in Section IV.

A limitation of this detection method is that it cannot detect the very highest vibrational
levels, because they can be photodissociated indirectly via off-resonant re-excitation by the PA
laser after they are formed.  This indirect photodissociation  process is discussed in detail in
Ref. \cite{Huang06a}.  There it was observed that only levels up to $v''=117$ can be observed,
even though the formation rate for high-$v''$ levels should have been very large.  In the present
work we confine our analysis to levels with $v''\leq116$, for which we estimate that losses due to
this process should be minor.

Another experimental limitation is that a small background signal level is observed even in the
absence of the PA laser, probably due to formation of high-$v''$ ground-state molecules by the MOT
laser itself.  After noting this issue in our first data set, we included in a second data set
explicit measurements of this background signal, taken by scanning the detection laser without the
presence of the PA laser.  This background level can be subtracted from each ion spectrum.
However, we note that the first data set cannot be so corrected, and also that the subtraction may
over-correct for the real background contribution, since the presence of the PA laser may cause
indirect photodissociation of the high-$v''$ molecules comprising the background, causing it to
diminish whenever the PA laser is present.  Thus we use in our analysis the average of the
uncorrected and corrected signals, and we take the difference between the two to be upper and
lower bounds when estimating the experimental uncertainties.

\section{Results and Analysis}
\label{sec:results}

\begin{figure}
\centering
\includegraphics[width=0.95\linewidth]{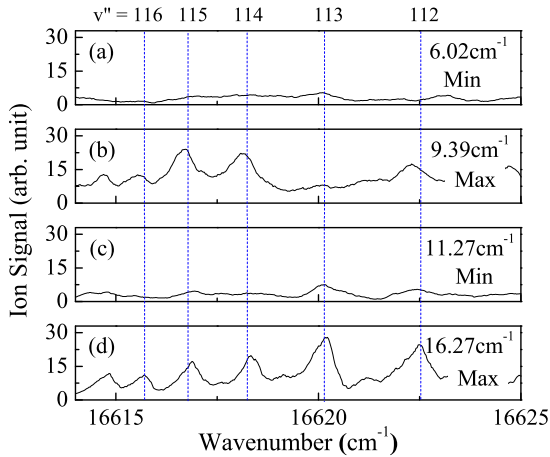}
\caption{(color online)
Comparison of experimental detection laser scans for PA detunings corresponding to the minimum and maximum effects of resonant coupling as predicted from the calculated curve in Fig. \ref{fig:FCFoscillations} Ionization signals for binding energies of (a) 6.02$\,$cm$^{-1}$ (Min) (b) 9.39$\,$cm$^{-1}$ (Max) (c) 11.27$\,$cm$^{-1}$ (Min), and (d)16.27$\,$cm$^{-1}$ (Max) are shown.  For easy comparison, the vibrational levels $v''$ are marked by vertical lines.}
\label{fig:rawdata}
\end{figure}

Two sets of experimental data were taken on different dates. Detection laser spectra were obtained
for each of nine different detunings $\Delta_{\textrm{PA}}$ of the PA laser, including the four
examples shown in Fig. \ref{fig:rawdata}. The data include all of the PA transitions in the range
$\Delta_{\textrm{PA}}=6-17$~cm$^{-1}$ that are sufficiently strong and free from overlaps with
unrelated transitions to allow reliable results. In this range, theory predicts two pairs of
maxima and minima for formation of molecules due to the resonant coupling phenomenon.  Assignment
of the vibrational levels $v''$=112-116 is based on our previous study of the formation of the
singlet molecules in the ground $X^1\Sigma_g^+$ state \cite{Huang06a}.

To show the enhancement of ultracold molecule formation in the $X~^1\Sigma_g^+$ ground state due
to resonant coupling, we compare the production of ground-state molecules for different PA laser
detunings.  However, the production rate involves both the efficiency of the PA step and the
efficiency of the spontaneous emission (SE) step, as is apparent in Fig. \ref{fig:potentials}.
Only the latter process is of interest to the present analysis.  Fortunately the effects of the
two processes can be distinguished by using trap loss measurements to determine the PA rate.
Production of excited molecules by PA results in trap loss with nearly 100\% probability, either
by formation of ground-state molecules or by decay back into the continuum with enough kinetic
energy gain to escape from the trap. When the PA laser is tuned to a resonance, the formation rate
$R$ of excited-state molecules is given by
\begin{equation}
R = {1 \over 2} L \frac{\Delta N}{N_o},
\end{equation}
where $L$ is the MOT loading rate, $N_o$ is the steady-state number of atoms in the MOT in the absence of the PA laser, and $\Delta N$ is the change in number due to the PA laser.  Therefore, the formation rate of excited-state molecules is proportional to the fractional trap loss with the assumption that the MOT parameters are fairly constant. Thus we can correct for variation in the PA rate at different PA laser detunings, by normalizing the ground-state molecule signal by the corresponding fractional trap loss.  In a typical data set, this fractional loss varies from 18\% to 27\% over our range of detunings.  As explained previously, we also sum together the signals (peak heights) from ground-state levels with $v''$=112-116 to average over nodal fluctuations in the Franck-Condon factors.

A small systematic error can occur because the detection laser beam intensity is not always
sufficient to saturate the first step of the detection process, the bound-bound $X~^1\Sigma_g^+
\rightarrow 2~^1\Sigma_u^+$ transition.  Therefore, we consider two limiting cases. First, we
assume that the transition is completely saturated.  Then the ionization signal should directly
reflect the Franck-Condon factors for the $0_u^+ \rightarrow X^1\Sigma_g^+$ transition.  Second,
we assume that there is no saturation at all. In this case the ionization signal must be
normalized by the Franck-Condon factors for the first detection step, $X~^1\Sigma_g^+ \rightarrow
2~^1\Sigma_u^+$.  Because the extent of saturation was not well-controlled during the
measurements, we use the average of these two limiting cases as our final experimental result.
The difference between them, which fortunately is small compared to the random experimental
scatter, is taken as the associated systematic uncertainty.

In Fig. \ref{fig:FCFoscillations}, the final experimental results for radiative decay rates to
levels with $v''=112-116$ are compared with theory.  Because the overall scale is arbitrary, we
choose for each data set a multiplicative scaling factor selected to minimize the standard
deviation between theory and experiment.  The typical systematic error bars shown in Fig.
\ref{fig:FCFoscillations} is estimated by taking the quadrature sum of the systematic
uncertainties due to background subtraction and incomplete detection laser saturation.  It does
not include the effects of random variations and day-to-day variations in the MOT, which are
apparent in the scatter between the two independent data sets.

The overall oscillatory behavior due to resonant coupling is very apparent in Fig.
\ref{fig:FCFoscillations}, and the positions of the minima and maxima coincide closely with the
theoretical predictions.  However, the ratio between the maximum and minimum molecule formation
rates is somewhat smaller than predicted by theory, about 5:1 as opposed to 20:1.  This may arise
from additional perturbations not taken into account in our two-state model, although it also may
be due in part to experimental issues such as the effects of weak overlapping transitions and the
saturation effects described above.  We also note that the relative populations of the individual
vibrational levels $v''$ exhibit variations different from the predicted Franck-Condon factors,
even though the broader distributions are well-predicted, indicating that the model is not quite
sufficient to predict the exact nodal positions in the vibrational wavefunctions.

\section{Conclusions}

We have presented clear theoretical and experimental evidence for
resonant enhancement in the photoassociative formation of ultracold
$^{85}$Rb$_2$ molecules in the ground $X\,^1\Sigma_g^+$ state.  The
basic mechanism, enhancement of radiative 0$_u^+ \rightarrow X$ decay
by resonant coupling of the 0$_u^+\,(1/2)$ and 0$_u^+\,(3/2)$ states,
is similar to that previously observed for Cs$_2$ \cite{DionPRL01}.  A
range of PA transitions was investigated with detunings of $6-17$
cm$^{-1}$ below the $5S+5P_{1/2}$ atomic asymptote.  A mapped grid
method was used to obtain vibrational wavefunctions to predict
relative molecule formation rates, and experiments were performed to
measure these rates.  The oscillatory behavior of the ground-state
molecule population has been revealed more clearly than in the prior
experimental work on Cs$_2$.  This was accomplished by normalizing the
experimental results to remove variations in PA rates and by
integrating the formation rates for several vibrational levels to
remove local nodal oscillations in the Franck-Condon factors.
Agreement between theory and experiment is quite good, except that the
observed contrast of the resonant enhancement phenomenon is somewhat
less than predicted.  The enhanced production of ground-state
molecules in the vibrational levels $v''=112-116$ should be quite
useful for future experimental studies. In particular, vibrational
levels with binding energies of several wavenumbers present a much
better starting point for additional Raman transitions to obtain
molecules in their vibronic
ground state than levels with sub-cm$^{-1}$ binding energies
\cite{KochPRA04}.

\begin{acknowledgments}
CPK is grateful to Tom Bergeman and Olivier Dulieu for making their potential energy curves, spin-orbit coupling functions and transition dipole functions available prior to publication.  Work in Berlin was supported by the Deutsche Forschungsgemeinschaft within the Emmy Noether programme.  Experimental work was supported by NSF and the University of Connecticut Research Foundation.
\end{acknowledgments}


\begin{thebibliography}{}
\bibitem{v=122}In order to accurately calculate the binding energy of the level
$v''=122$, one would need to take the hyperfine interaction into account.
\end{thebibliography}
\end{document}